\documentclass[twocolumn,superscriptaddress,floatfix,prl,aps,showpacs]{revtex4}
\usepackage{graphicx}
\usepackage{latexsym}
\usepackage{amsmath}
\begin{document}
\bibliographystyle{apsrev}
\title{Universal alternating order around impurities in antiferromagnets}
\author{Sebastian Eggert}
\affiliation{Dept.~of Physics, Univ.~of Kaiserslautern,  D-67663 Kaiserslautern, Germany}
\author{Olav F. Sylju\aa sen}
\affiliation{NORDITA, Blegdamsvej 17, DK-2100 Copenhagen \O, Denmark}
\author{Fabrizio Anfuso}
\affiliation{ Theoretical Physics, 
University of K\"oln, D-50937 K\"oln,  Germany }
\author{Markus Andres}
\affiliation{Dept.~of Physics, Univ.~of Kaiserslautern,  D-67663 Kaiserslautern, Germany}

\begin{abstract}
The study of impurities in antiferromagnets is of considerable interest in
condensed matter physics.  In this paper we address the elementary question of 
the effect of vacancies on the orientation of the 
surrounding magnetic moments in an antiferromagnet.  In the presence of a 
magnetic field, alternating magnetic moments are induced, which can be described 
by a universal expression that is valid in any ordered antiferromagnet and 
turns out to be independent of temperature over a large range.  The universality is 
not destroyed by quantum fluctuation, which is demonstrated 
by quantum Monte Carlo simulations in the two-dimensional Heisenberg antiferromagnet. 
Physical predictions for finite doping are made, which are relevant 
for experiments probing Knight shifts and the order parameter.
\end{abstract}
\maketitle

The intentional doping of antiferromagnetic
materials has become a useful tool in order to study the
complicated physics in the context of high temperature superconductivity and
quantum magnetism\cite{Vajk,Sachdev,Mahajan,julien,bobroff,ouazi}.
Large alternating magnetic moments
around static non-magnetic impurities are observed in Knight shift experiments
when a uniform field is applied\cite{Mahajan,julien,bobroff,ouazi}.
Theoretical studies have shown 
that vacancies in low-dimensional antiferromagnetic backgrounds 
give rise to locally enhanced antiferromagnetic 
correlations\cite{Bulut,eggert,Martins,Sandvik2,anfuso,Bulut1,rommer,eggert2},
which strongly depend on the microscopic model and temperature in the low dimensional
models.

\begin{figure}
\includegraphics[width=0.48\textwidth]{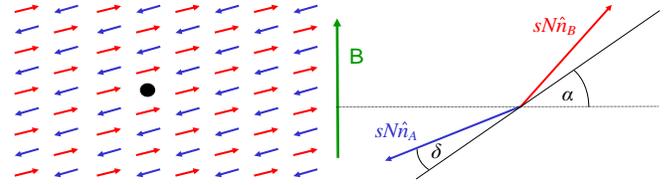}
\caption{(Color online)
Schematic illustration of the effect of a vacancy in an ordered antiferromagnet.
The spins ''cant'' with an angle $\delta$ towards the field corresponding to a small 
{\it uniform} 
magnetization. Due to the broken sub-lattice symmetry  
the order may be ''tilted'' by an angle $\alpha$ relative to the plane normal to the field
corresponding to an induced {\it alternating} magnetization around the impurity.
}
\label{scheme}
\end{figure}

In this work, we 
show that in generic {\it ordered} antiferromagnets 
the alternating local moments in the vicinity of vacancies
can be quantitatively described by a universal expression which only depends on the 
field $B$, but is surprisingly independent of temperature, quantum fluctuations, and
microscopic details.
The mechanism which gives rise to the alternating moments is a local tilting of
the order parameter due to the broken sub-lattice symmetry by impurities.
In contrast to the pure sample, where the 
order parameter is always confined in the plane normal to the field, 
a large alternating order {\it parallel} to the field is induced as 
schematically depicted in Fig.~\ref{scheme}.
The calculations agree remarkably well with quantum Monte Carlo (QMC) simulations without any adjustable parameters even in 
two dimensions $D=2$, where quantum fluctuations are strongest.

The typical antiferromagnetic Hamiltonian
\begin{equation}
H=J\sum_{\langle i,j \rangle}\mathbf{S}_{i}\cdot \mathbf{S}_{j} - \sum_j B 
{S}_{j}^z, 
\label{ham}
\end{equation}
describes the magnetic behavior realistically even for rather 
complex materials despite its simplicity.  We consider systems with bipartite lattices
of dimension $D \geq 2$, where the sum in Eq.~(\ref{ham}) runs over nearest neighbor sites.
Generically, the dominant interaction $J>0$ comes from the Coulomb forces via the 
exchange mechanism and is therefore isotropic.  The rotational symmetry is broken by 
applied and crystal fields ${B}$ in units of $g\mu_B$, which are typically
small compared to the interaction $B < J$.  The direction of the field defines the z-axis
of our coordinate system, which does not need to coincide with any of the lattice directions.
For bipartite lattices of dimension $D \geq 2$ the model system (\ref{ham}) is known to 
have finite N\'eel order at sufficiently low temperatures for both 
quantum and classical spins $\mathbf{S}_{i}$ of any size $s$\cite{footnote}.  
The ordered state remains stable over a large range of 
perturbations by impurities and frustrating interactions.

In order to obtain an 
intuitive picture of the physical behavior, let us first consider 
a highly simplified model of the Hamiltonian (\ref{ham}).  The long-range order 
spreads over the entire sample, so it might be justified to describe all ordered spins on 
one sub-lattice $A$ by a common direction 
$\hat{n}_A = (\sin \theta_A \, \sin \phi_A, \sin \theta_A \, \cos \phi_A, \cos \theta_A)$ 
and analogously for sub-lattice $B$.  
In this case, the interaction energy is always minimized by a relative
azimuthal angle $\phi_A - \phi_B = \pi$, so that the effective energy is given 
just in terms of the polar angles 
\begin{eqnarray}
E_{\rm eff}  & = &  J z N s^2 \hat n_A \cdot\hat n_B - s N B (n_A^z+n_B^z) 
\nonumber \\
& = &  J z N s^2 \cos (\theta_A+\theta_B) - s N B ( \cos \theta_A + \cos \theta_B) \nonumber \\
& = &  J z N s^2 (2 \sin^2 \delta - 1)- 2 s N B \cos \alpha \sin \delta \label{Eeff} 
\end{eqnarray}
where $z$ is the number of nearest neighbors and 
$N$ is the total number of sub-lattice sites in the sample.
The angle $\delta= (\pi - \theta_A- \theta_B)/2$ corresponds to  
a uniform ''canting''  of all spins on both sub-lattices towards an applied magnetic field
as shown in Fig.~\ref{scheme}. 
The angle $\alpha=(\theta_B- \theta_A)/2$ measures the 
''tilt'' of the antiferromagnetic order 
relative to the plane that is normal to the field.

Below saturation $B < 2 s z J$ the energy is minimized by setting
$\sin \delta = \frac{B}{2 s z J}\cos\alpha$ which gives an effective low energy 
description for $\alpha$
\begin{equation}
E_{\rm eff}(\alpha) = \frac{N B^2}{2 z J}\sin^2\alpha +E_0
\end{equation}
where $E_0 = -J N s^2 z - NB^2/2 z J$.  The physical interpretation of this 
simple model is textbook knowledge\cite{AM}:
All spins align slightly towards the magnetic field 
$m = s \langle \sin\delta\rangle = B \chi_\perp \cos\alpha$ 
with a susceptibility $\chi_\perp \sim  1/2zJ$
that is largest 
when the magnetic field is perpendicular to the order
and therefore there is a small energy gain 
for the N\'eel order to be in the plane normal to $B$ (i.e.~$\alpha=0$).  
Since the energy gain is 
small, the order may point in another direction in realistic materials where 
the sub-lattice symmetry is broken.  A common source of sub-lattice symmetry breaking is 
disorder and impurities which is the topic of this paper.

Let us first consider a single vacancy in the framework of the simple model above by 
reducing the size of the corresponding sub-lattice vector by one spin
$\mathbf{N}_A = s (N-1) \hat{n}_A$.  Starting from Eq.~(\ref{Eeff}) the 
susceptibility 
for $\delta$ remains the same for large $N$. However, the size
of the two sub-lattice spins is not equally large and therefore a net coupling to 
the field remains in the effective energy as a function of the z-component of the 
alternating order $n_z= \sin \alpha$
\begin{equation}
E_{\rm eff}(n_z) = {N B^2 \chi_\perp }n_z^2 -s B n_z + E_0
\label{Ealpha}
\end{equation}
where we have also used that the dependence on $\delta$ 
is small and irrelevant in the direct coupling term.
Even though $N$ is large, the second term will ensure that the expectation value of
the impurity induced alternating order along the field $n_z$ is always non-zero
\begin{eqnarray}
\langle n_z \rangle
 & = &   
   \frac{1}{Z} \int_{-1}^{1} dn_z \ 
e^{-\beta E_{\rm eff}(n_z)} n_z \nonumber \\
 & = &   \frac{s}{B N \chi_\perp}\left(\frac{1}{2} - 
\frac{e^{-B^2 N \chi_\perp \beta}}{\int_{-1}^{1}e^{-x^2 B^2 N \chi_\perp \beta}dx}
\right)\label{malt1}
\end{eqnarray}
where we have assumed the thermodynamic limit $N \gg \beta J$ ($\beta = 1/k_BT$).
In the limit of large and small fields, respectively,  we find
\begin{equation}
\langle n_z \rangle =  \left\{ \begin{array}{lcl}
   {s B}/{3T}& \phantom{nnn} & {\rm for} \ \   N \chi_\perp B^2 \ll T  \\
  ~ & ~ & ~ \\ 
   {s}/{2 N B \chi_\perp} & \phantom{nnn} & {\rm for} \ \ N \chi_\perp B^2 \gg T \\
\end{array}
\right.
\label{malt2}
\end{equation}

In the first case of very small fields, the alternating response to a uniform field
is described by 
a classical susceptibility, which also directly follows from Eq.~(\ref{Ealpha}) 
if only the second term is kept (i.e.~$\chi_\perp B \to 0$).
Therefore a tilting of the order parameter of order $\alpha \sim B/T$ is 
expected which is larger than the uniform canting $\delta \sim B/J$
in the ordered phase $T \ll J$.  By QMC simulations it was shown that a corresponding 
alternating order is induced throughout the lattice by a single vacancy in the limit 
of linear response\cite{anfuso}, which is consistent with the assumption that 
$\alpha$ describes the tilting of all spins.
The corresponding impurity susceptibility
is given by a classical Curie behavior $s^2/3T$, 
as first predicted in Ref.~[\onlinecite{Sachdev}] and confirmed by 
QMC simulations in Ref.~[\onlinecite{Sandvik}] 
in the limit of linear response $N \chi_\perp B^2 \ll T$. 
This limit is only relevant in the case where the domain size $N$ is 
restricted by disorder or boundaries.

However, if $N$ is macroscopic, the limit
$N \chi_\perp B^2 \gg T$ is already reached for any naturally 
occurring background field, so that 
the second case in Eq.~(\ref{malt2}) 
is the more interesting
limit for the description of realistic impurity effects.  
The induced alternating magnetization decreases with increasing field and
the behavior is independent of temperature since corrections 
from the second term in Eq.~(\ref{malt1}) are exponentially 
small in the macroscopically 
large scaling variable $N \chi_\perp B^2 \beta$.  This remarkable 
behavior will survive even in more refined models and give rise to a universal 
temperature independent description as we will see.

In order to make the model more realistic, the angle $\alpha$
can be interpreted as a {\it local} tilting that is dependent on position in 
order to reflect the fact that the first term in Eq.~(\ref{Ealpha})
is an effective potential that acts on all spins in the lattice, 
while the second term arises from the vacancy locally at the origin.
There is an energy cost to change the direction of the order parameter from one 
lattice site to a neighboring lattice site corresponding to 
the so-called spin-stiffness $\rho_s$, so that 
Eq.~(\ref{Ealpha}) has to be generalized to an 
energy functional for $n_z$ 
\begin{eqnarray} 
E\left[n_z(\mathbf{r})\right]  &  = & 
\int d^D \mathbf{r} \ \left(\frac{\rho_s}{2} \left( \nabla n_z(\mathbf{r}) \right)^2 
+\frac{\chi_\perp}{2} B^2 n_z^2(\mathbf{r}) \right) 
\nonumber \\ ~ &~  &  
-s B n_z(0),
\label{funct}
\end{eqnarray}
where we have replaced the sum over both sub-lattices by an integral for convenience.
The energy density in the first term is reminiscent of the non-linear 
sigma model\cite{sigma3,sigma4}, but only for one component 
and without the imaginary time direction describing the quantum fluctuations.

In order to calculate the expectation value of 
$\langle n_z(\mathbf{r_0})\rangle$ at any position 
$\mathbf{r_0}$, it is useful to define 
a generating partition function 
\begin{equation}
Z_\gamma = \int {\cal D}[n_z(\mathbf{r})] \exp\left(-\beta E\left[n_z(\mathbf{r})\right] + 
\gamma n_z(\mathbf{r_0})\right). \label{Z}
\end{equation}
The expectation value is then given by the logarithmic derivative
\begin{equation}
\langle n_z(\mathbf{r_0}) \rangle = \partial_\gamma \ln Z_\gamma |_{\gamma=0}.
\label{theta}
\end{equation}
In momentum space the partition function becomes
\begin{equation}
Z_\gamma = \int {\cal D}[n_z(\mathbf{q})] e^{\left[ \int d^D\mathbf{q}
\left(-{\beta}  E_\mathbf{q} 
|n_z(\mathbf{q})|^2 +I_\mathbf{q}(\gamma) 
n_z(\mathbf{q})\right) \right]}
\label{Zq}
\end{equation}
where 
\begin{eqnarray}
E_\mathbf{q} & = & \left( \rho_s q^2+\chi_\perp B^2 \right)/2 \\
I_\mathbf{q}(\gamma) & = & \left(\beta B s + \gamma \cos \mathbf{q}\cdot 
\mathbf{r_0}\right)/(2\pi)^{D/2}. \label{J}
\end{eqnarray}
The expectation value is therefore 
\begin{eqnarray}
 \langle n_z(\mathbf{r_0}) \rangle  & = &   \partial_\gamma
\ln Z_\gamma|_{\gamma=0}  \nonumber \\
& = &    \partial_\gamma \int d^D\mathbf{q} \ln \int d n_z\  
e^{-{\beta}
E_\mathbf{q} n_z^2 +I_\mathbf{q}(\gamma) n_z}
|_{\gamma=0}  \nonumber \\
& = &  \left. \partial_\gamma \int \frac{d^D\mathbf{q}}{(2 \pi)^{D/2}} \frac{I^2_\mathbf{q}(\gamma) 
}{2 \beta (\rho_s q^2 +\chi_\perp B^2)} \right|_{\gamma=0} \nonumber\\
& = &  \int \frac{d^D\mathbf{q}}{(2 \pi)^D} \frac{B s \cos (\mathbf{q}\cdot
\mathbf{r_0})
}{(\rho_s q^2 +\chi_\perp B^2)} \label{universal}\\
& =  &  \left\{ \begin{array}{lcl}
\frac{s \ B}{2 \pi \rho_s}K_0\left( \frac{B }{c}r \right) & \phantom{nnn} & D=2  \\
     ~ & ~ & ~ \\
      \frac{s \ B}{4 \pi \rho_s r} e^{-Br/c} & \phantom{nnn} & D=3  \\
      \end{array}
      \right.
      \label{bessel}
\end{eqnarray}
where $c=\sqrt{\rho_s/\chi_\perp}$ is known as the spin-wave velocity and $K_0$
is the modified Bessel function.
This result is remarkable in two ways:  first of all it turns out to be 
completely independent of 
temperature and secondly it is independent of the underlying detailed 
microscopic model.  Therefore, the formula in Eq.~(\ref{universal}) can be taken 
as a universal description for all antiferromagnets in the ordered phase.  
Variations in the lattice structure, frustration, quantum effects, and the detailed 
microscopic parameters 
only renormalize the spin stiffness $\rho_s$ and the uniform susceptibility $\chi_\perp$, 
but not the functional behavior in Eq.~(\ref{bessel}).  
For spins close to the vacancy $r \agt 1$ the tilting $n_z \sim s B/4 \pi \rho_s$ 
remains typically less than saturation, but 
larger than the uniform canting $\alpha > \delta$, so that 
spins on the same sublattice as the vacancy tend to align {\it against} the field.

It can be checked that the functions in Eq.~(\ref{bessel}) are solutions of the 
diffusion equation 
\begin{equation}
B^2 \chi_\perp n_z = \rho_s \nabla^2 n_z
\end{equation}
that also follows from minimizing the energy 
functional Eq.~(\ref{funct}). 
In lattices where the spin-stiffness is not isotropic, the result can be generalized
by taking $\rho_s$ as an anisotropic diffusion coefficient.

As a concrete example, we will now consider 
the spin-1/2 Heisenberg model on a 2D square lattice, which is
possibly the most studied antiferromagnetic model, since it 
has received much attention in connection with 
high temperature superconductivity, but also because it is an interesting 
case where quantum fluctuations strongly compete with N\'eel order.

\begin{figure}
\includegraphics[width=0.48\textwidth]{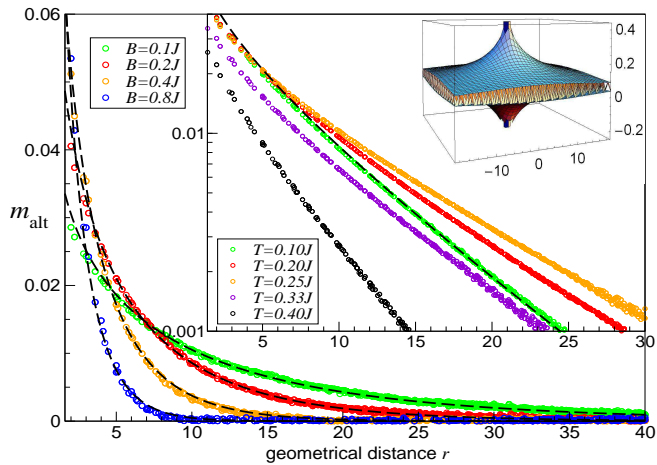}
\caption{(Color online) The alternating response $m_{\rm alt}$ as a function of 
geometrical distance $r$ from a vacancy in a 2DHAF
at different fields and $T=0.025J$ from QMC simulations compared to the universal 
theoretical
prediction  $m_{\rm max} n_z$ in Eq.~(\ref{2d}) {\it without any adjustable parameters} (dashed black lines).
Inset:  $m_{\rm alt}$ at $B=0.2 J$.  
Even at higher $T=0.1J$ no deviations from Eq.~(\ref{2d}) 
can be seen on a logarithmic scale.  At still higher $T$
the induced order first {\it increases} for $0.3J \agt T \agt 0.2J$ and then 
decreases again for $T \agt 0.3J$.
Inner inset: Magnetic moments at $B=0.1J$ and $T=0.1J$ alternating between the 
extrapolated amplitudes on even and odd sublattices.}

\label{plots}
\end{figure}

In Monte Carlo simulations we have used the stochastic series expansion with 
directed loop updates\cite{QMC}
in order to extract 
the magnetic moments $\langle m_z\rangle$ around a single vacancy in small magnetic fields 
in the ordered phase $\xi(T) \gg L$ (here $L=128$) as shown in the inner 
inset of Fig.~\ref{plots}.
In the plane perpendicular to the field the order is fluctuating, so that
$\langle m_x\rangle =\langle m_y \rangle = 0$.
For the moments parallel to the field, we expect to find a large
staggered magnetization according to Eq.~(\ref{bessel})  
\begin{equation}
\langle m_z(\mathbf{r})\rangle = (-1)^r m_{\rm max}
\frac{s \ B}{2 \pi \rho_s}K_0\left( \frac{B }{c}r \right)
\label{2d}
\end{equation}
in addition to the less interesting small uniform canting $\delta$.
Here $m_{\rm max} \approx 0.308$ is the 
maximum order in the 2D Heisenberg model which is reduced from $s=1/2$ due to 
quantum fluctuations. In fact, there are no adjustable parameters in Eq.~(\ref{2d})
since all relevant parameters have long been established to high precision 
by independent methods\cite{igarashi}: 
\begin{eqnarray}
m_{\rm max} \approx 0.308, \  & \ & \  \rho_s \approx 0.18J, \nonumber \\
 \chi_\perp \approx 0.065/J, \ & \ &  \ c = \sqrt{\rho_s/\chi_\perp} \approx 1.67J.
\label{constants}
\end{eqnarray}

By extrapolating the numerical data for $m_z(\mathbf{r})$ on the even and
the odd sub-lattice 
separately and taking half the difference,
we extracted the staggered magnetization $m_{\rm alt}(\mathbf{r})$ around a vacancy.
The resulting alternating amplitude $m_{\rm alt}$ is completely isotropic and can be 
plotted as a function of the 
geometrical distance $r=| \mathbf{r}|$ only as shown in Fig.~\ref{plots} for 
different fields and temperatures.  While the size of $m_{\rm alt}$
is proportional to the field, the
drop-off is shortened for higher fields so that the integrated amplitude 
decreases with increasing field as also reflected in the simple model of Eq.~(\ref{malt1}).
The agreement with Eq.~(\ref{2d}) is remarkably good even on a logarithmic scale
and for widely different fields and temperatures, which 
we take as confirmation for the 
general validity of the result in Eq.~(\ref{universal}).
Since there were no adjustable parameters, we
conclude that the quantitative predictive power for static expectation values
of the hydrodynamic model in Eq.~(\ref{funct}) 
is not changed by quantum fluctuations. 
From a theoretical point of view this means that the renormalized classical 
model\cite{sigma3,sigma4} can be taken 
for quantitative calculations anywhere in the antiferromagnetic phase, while
microscopic details only affect the values of the constants in Eq.~(\ref{constants}).
In particular, close to a critical point $\rho_s$ and $m_{\rm max}$
become vanishingly small, but the model remains valid.

A breakdown of the universal formula in Eq.~(\ref{universal}) must occur at 
the transition temperature to the disordered phase.
In the 2D simulations we find indeed that any temperature dependence is
exponentially supressed until the Kosterlitz-Thouless temperature is approached\cite{KT}
$T_{KT} \sim 0.2J$.  However, the induced alternating magnetization is
surprisingly {\it enhanced} by increasing temperature near $T_{KT}$ as shown in the
inset of Fig.~\ref{plots}.  Only at still higher temperatures $T\agt 0.3J$
the induced order is finally reduced as expected,
leading to a non-monotonic behavior with temperature.
While we have no explanation of this exotic effect in terms of our model, 
we hope that future works on this topic may uncover this mystery.

Finally, we would like to generalize our results to finite impurity 
concentrations $\rho$.
For higher fields/small concentrations $\rho < (B/c)^D$ the impurities are
sufficiently far apart to be treated independently (dilute limit).  In this case, 
the above conclusions are unchanged and the magnetic order is tilted locally in
the vicinity of each vacancy.   At smaller fields/larger concentrations 
$\rho > (B/c)^D$
the impurities become correlated and enhance/annihilate the tilting effect
depending if they are on the same/opposite sub-lattices\cite{anfuso}.
In this disorder limit all impurities become correlated and the 
tilting is again nearly uniform throughout the lattice, with an effective 
total impurity strength that is given by the difference of the vacant sites
on each sub-lattice $| N_A -N_B| \sim \sqrt{\rho N}$ in a domain of size $N$.
In this case the simple model in Eq.~(\ref{Ealpha}) remains valid with the 
effective size of the spin $s$ in the last term replaced by $s\sqrt{\rho N}$. 
The average universal tilt in Eq.~(\ref{malt2}) throughout the domain
is then given by $m_{\rm max} s \sqrt{\rho}/2 \sqrt{N} B \chi_\perp$.

In conclusion, we have analyzed the induced alternating magnetization around 
vacancies in ordered antiferromagnets in quantitative detail.
Large alternating moments are induced parallel to the field, which 
corresponds to a tilting of the order parameter.
The induced order decays with distance
at a rate that is independent of temperature and inversely proportional
to the field $c/B$.  
From a theoretical point of view we have demonstrated that the renormalized classical 
description gives an intuitive insight into the mechanism on how the alternating magnetization
arises.  At the same time, the theory gives good quantitative agreement with 
QMC 
simulations even in 2D where quantum fluctuations are large.
At large impurity densities $\rho>(B/c)^D$ the
impurities become correlated and give rise to a tilt of the 
order throughout the sample
towards the field direction.  The predicted effects can be observed by 
Knight shift experiments 
like NMR and $\mu$SR, or by investigating the order directly via 
magnetic neutron scattering.  
Numerically we find an {\it enhancement} of the 
induced order near the Kosterlitz-Thouless 
transition $T_{KT} \agt 0.2J$ which is counter-intuitive and calls for further
investigation.

\begin{acknowledgments}
We are grateful to Stellan \"Ostlund for discussions which inspired us to 
investigate this topic. The collaboration was supported by the Nordforsk Network on 
Low Dimensional Physics.
\end{acknowledgments}

\end{document}